\documentclass [aps,prl,twocolumn,showpacs,nofootinbib,superscriptaddress]{revtex4}
\usepackage{verbatim}
\usepackage{slashed}
\usepackage{epsfig}
\usepackage{axodraw}
\usepackage{graphicx}
\usepackage{amsmath}
\usepackage{mathrsfs}
\usepackage{bm}

\begin{document}


\title{ QCD prediction for the non-$D\bar D$ annihilation decay of $\psi(3770)$}


\author{Zhi-Guo He}
\affiliation{Department of Physics and State Key Laboratory of
Nuclear Physics and Technology, Peking University,
 Beijing 100871, China}
\author{Ying Fan}
\affiliation{Department of Physics and State Key Laboratory of
Nuclear Physics and Technology, Peking University,
 Beijing 100871, China}
\author{Kuang-Ta Chao}
\affiliation{Department of Physics and State Key Laboratory of
Nuclear Physics and Technology, Peking University,
 Beijing 100871, China}




\begin{abstract}
To clarify the marked difference between BES and CLEO measurements
on the non-$D\bar D$ decays of the $\psi(3770)$, a
$1^{3}D_{1}$-dominated charmonium, we calculate the annihilation
decay of $\psi(3770)$ in NRQCD. By introducing the color-octet
contributions, the results are free from infrared divergences. The
color-octet matrix elements are estimated by solving the evolution
equations. The S-D mixing effect is found to be very small. With
$m_{c}=1.5\pm0.1\textrm{GeV}$ our result is
$\Gamma(\psi(3770)\rightarrow \textrm{light~
hadrons})=467^{-187}_{+338}\textrm{KeV}$. For $m_c=1.4$~GeV,
together with the observed hadronic transitions and E1 transitions,
the non-$D\bar D$ decay branching ratio of $\psi(3770)$ could reach
about 5\%. Our results do not favor either of the results of BES and
CLEO collaborations, and further experimental tests are urged.
\end{abstract}

\pacs{12.38.Bx, 12.39.Jh, 13.20.Gd,}

\maketitle


Heavy quarkonia decays play an important role in understanding
quantum chromodynamics (QCD)\cite{yellow}. These include not only
the determination of the running strong coupling constant
$\alpha_{s}$ from S-wave decays $J/\psi\to ggg$ and $\Upsilon\to
ggg$, but also the study of factorization from the P-wave
annihilation decays, where appear infrared (IR) divergences in
$^1P_1\to ggg$ and $^3P_J\to gq\bar q $ \cite{BBL5,BBL6}. A
traditional way to treat the IR divergences was to use the quark
binding energy or the gluon momentum as cutoff to estimate these IR
divergences, but this is model dependent and breaks factorization of
short and long distance processes. In~\cite{BBL1}, a new
factorization scheme was proposed
to absorb the IR logarithms by  new non-perturbative parameters, the
color octet matrix elements. Based on the non-relativistic nature of
heavy quarkonia, an effective theory, Non-Relativistic QCD (NRQCD)
was developed~\cite{BBL2}, in which the inclusive annihilation
decays can be calculated in a systematic way by double expansions in
terms of $\alpha_{s}$ and $v$, the relative velocity of quarks in
heavy quarkonium. In \cite{Huang1,Huang2,Huang3}, the authors
calculated QCD radiative corrections to the light hadron (LH) decays
of P-wave charmonium in NRQCD, and showed explicitly the cancelation
of infrared divergences at the next to leading order (NLO).
In \cite{maltoni} a more complete and precise NLO calculation for
the P-wave decay perturbative coefficients in NRQCD is given (see
also \cite{Maltoni00}). At NLO in $\alpha_{s}$, the NRQCD
predictions for the relative decay rates of $\chi_{cJ}\to LH$ are
consistent with more updated data (see Chapter 4 of \cite{yellow}).
Moreover, the relativistic corrections of S and P-wave
electromagnetic quarkonium decays have been given at order
$v^{7}$\cite{brambilla}. As for the D-wave, in
\cite{L.Berg,G.Belanger} calculations of $^{3}D_{J}\to ggg$ decays
were given but suffered from IR divergences; while in \cite{Huang3}
only the leading order (LO) color-octet contribution to
$^{3}D_{J}\to LH$ was given. So for the D-wave a complete
calculation for the IR cancelation and radiative correction in NRQCD
is apparently needed.

Phenomenologically, for the $^{3}D_{1}$ ($J^{PC}=1^{--}$) charmonium
state $\psi(3770)$, there is a long-standing puzzle in its
non-$D\bar D$ decays that the $\psi(3770)$ might have substantial
decays not into $D^0\bar D^0$ and $D^+D^-$.
BES earlier reported two results based on different analysis
methods: $\textrm{Br}(\psi(3770)\rightarrow
\textrm{non-}D\bar{D})=(14.5\pm1.7\pm5.8)\%$\cite{BES2}, and
$\textrm{Br}(\psi(3770)\rightarrow
\textrm{non-}D\bar{D})=(16.4\pm7.3\pm4.2)\%$\cite{BES3}. In
contrast, CLEO\cite{CLEO} measured the cross section
$\sigma(e^{+}e^{-}\rightarrow\psi(3770)\rightarrow
\textrm{non-}D\bar{D})=-0.01\pm0.08^{+0.41}_{-0.30} \textrm{nb}$.
Very recently, with the first direct measurement on the non-$D\bar
D$ decay, BES gives
$\sigma(e^{+}e^{-}\rightarrow\psi(3770)\rightarrow
\textrm{non-}D\bar D)=(0.95\pm0.35\pm0.29)$ nb and
$\textrm{Br}(\psi(3770)\rightarrow\textrm{non-}D\bar{D})=(13.4\pm5.0\pm3.6)\%$\cite{BES4}.
Evidently, the two collaborations give very different results of the
non-$D\bar D$ decay of $\psi(3770)$. Meanwhile, a number of
experiments to search for the exclusive hadronic non-$D\bar D$
decays of $\psi(3770)$ have been done by BES\cite{BES5} and
CLEO\cite{CLEO1}, but no significant signals are found.

At least two kinds of non-$D\bar D$ decays of $\psi(3770)$ have been
observed. The hadronic transitions  $\psi(3770)\to \pi^+\pi^-J/\psi$
was first observed by BES with a branching ratio of $(0.34\pm
0.14\pm 0.09)\%$\cite{BES6}, and  was later confirmed by CLEO with a
somewhat smaller branching ratio Br$(\psi(3770)\to
\pi^+\pi^-J/\psi)=(0.189\pm 0.020\pm 0.020)\%$\cite{CLEO2}, and the
$\pi^0\pi^0 J/\psi$ and $\eta J/\psi$ modes were also seen with each
having a branching ratio of about one half of that of
$\pi^+\pi^-J/\psi$~\cite{CLEO2}. These results are within the range
of theoretical predictions based on the QCD multipole expansion for
hadronic transitions~\cite{Kuang}. With the total width of
$23.0\pm2.7$~Mev for $\psi(3770)$~\cite{PDG}, the width of all
hadronic transitions is about $100-150$ KeV. Another kind of
non-$D\bar D$ decays of $\psi(3770)$ are the E1 transitions
$\psi(3770)\rightarrow \gamma+\chi_{cJ}$ (J=0,1,2), and their widths
are measured by CLEO to be $172\pm 30, 70\pm 17, <21$~KeV for
J=0,1,2 respectively~\cite{CLEO3}, which are in good agreement with
predicted values 199, 72, 3.0~KeV in a QCD-inspired potential model
calculation with relativistic corrections~\cite{ding} (see
also~\cite{rosner,barnes}). The width of all E1 transitions
$\psi(3770)\rightarrow \gamma+\chi_{cJ}$ (J=0,1,2) is about
$250\pm50$KeV. The above mentioned hadronic and E1 transitions only
contribute 350-400~KeV and 1.5-1.8\% to the non-$D\bar D$ decay
width and branching ratio of $\psi(3770)$.



To clarify the puzzle of $\psi(3770)$ $\textrm{non-}D\bar D$ decay,
in this letter we will give a complete infrared safe NLO QCD
corrections to the annihilation decay rate of the $\psi(3770)$ in
the framework of NRQCD. Since $v^{2}\sim \alpha_{s}(m_{c})\approx
0.3$ in charmonium, the relativistic corrections are also important
and  should be considered in the future work.

The  $\psi(3770)$ can be viewed as a $1^{3}D_{1}$ dominated state
with a small admixture of $2^{3}S_{1}$, and expressed as (see e.g.
\cite{ding,rosner})
\begin{eqnarray}
|\psi(3770)\rangle=\cos\theta|1^3D_{1}\rangle+\sin\theta|2^3S_{1}\rangle,
\nonumber\\
|\psi(3686)\rangle=-\sin\theta|1^3D_{1}\rangle+\cos\theta|2^3S_{1}\rangle,
\end{eqnarray}
where $\theta$ is the S-D mixing angle  and it is about
$(12\pm2)^\circ$ by fitting the leptonic decay widths of
$\psi(3770)$ and $\psi(3686)$. Then the LH decay width of
$\psi(3770)$ is
\begin{eqnarray} \label{eq2}
&&\Gamma(\psi(3770)\rightarrow LH)=\cos^2\theta
\Gamma(1^3D_{1}\rightarrow LH)+\nonumber\\ &&\sin^2\theta
\Gamma(2^3S_{1}\rightarrow LH)+IF,
\end{eqnarray}
where $IF$ stands for the S-D interference term. The calculation of
S-wave decay at order $\alpha_{s}^{3}$ and leading order in $v^2$ is
trivial, and it gives
\begin{equation}
\Gamma(2^3S_{1}\rightarrow
LH)=\frac{|R_{2S}(0)|^2}{4\pi}\frac{40\alpha_{s}^{3}(\pi^{2}-9)}{81m_c^2},
\end{equation}
where $R_{2S}(0)$ is the $2^3S_{1}$ wave function at the origin. The
S-D interference term $IF$ in Eq.(2) is infrared finite at leading
order in $v^2$ and $\alpha_{s}$, and can be obtained by combining
the $1^3D_{1}\to 3g$ with $2^3S_{1}\to 3g$ amplitudes
\begin{eqnarray}
IF=2\sin\theta\cos\theta\frac{5(-240+71\pi^2)\alpha_{s}^{3}}{324m_{c}^4}
\frac{R_{2S}(0)}{\sqrt{4\pi}}\sqrt{\frac{1}{8\pi}}R_{1D}''(0),
\end{eqnarray}
where $R''_{D}(0)$ is the second derivative of the $1^3D_{1}$ wave
function at the origin.

We now proceed with the calculation of the main part, the D-wave
quarkonium LH decay. In NRQCD, the inclusive annihilation decay of
$^{3}D_{1}$ at leading order in $v^2$ is factorized as
\begin{widetext}
\begin{equation}\label{eq1}
\Gamma(^{3}D_{J}\rightarrow LH)=2\textrm{Im}
f(^{3}D_{J}^{[1]})H_{D1}+\sum_{J=0} ^{2}2\textrm{Im}
f(^{3}P_{J}^{[8]})H_{P8J}+2\textrm{Im}
f(^{3}S_{1}^{[8]})H_{S8}+2\textrm{Im} f(^{3}S_{1}^{[1]})H_{S1},
\end{equation}
\end{widetext}
where $\textrm{Im}f(n)$ is the imaginary part of the $Q\bar
Q\rightarrow Q\bar Q$ scattering amplitude,  and can be calculated
perturbatively . And the corresponding non-perturbative matrix
elements are
\begin{eqnarray}
H_{D1}=\frac{\langle H|
\mathcal{O}_{1}(^{3}D_{1})|H\rangle}{m_{c}^{6}},
H_{P8J}=\frac{\langle H|
\mathcal{O}_{8}(^{3}P_{J})|H\rangle}{m_{c}^{4}},\nonumber\\
H_{S8}=\frac{\langle H|
\mathcal{O}_{8}(^{3}S_{1})|H\rangle}{m_{c}^{2}}
,H_{S1}=\frac{\langle H|
\mathcal{O}_{1}(^{3}S_{1})|H\rangle}{m_{c}^{2}},
\end{eqnarray}
where $H$ is $\psi(1^3D_{1})$. Those four-fermion operators of
S-wave and P-wave are defined in \cite{BBL2}, and here we only give
the definition of the D-wave four-fermion operator
(the normalization of the color singlet four-fermion operators agree
with those in \cite{maltoni}):
\begin{align}
\mathcal{O}_{1}(^{3}D_{1})=\frac{3}{10N_{c}}\psi^{\dagger}\boldsymbol{T}^{i}\chi
\chi^{\dagger}\boldsymbol{T}^{i}\psi,
\end{align}
where
$\boldsymbol{T}^{i}=\boldsymbol{\sigma}^{j}\boldsymbol{S}^{ij}$ and
$\boldsymbol{S}^{kl}=(\frac{-i}{2})^{2}
(\overleftrightarrow{D}^{i}\overleftrightarrow{D}^{j}-\frac{1}{3}\overleftrightarrow{D}^{2}\delta^{ij})$.

We calculate the short distance coefficients at order
$\alpha_{s}^3$,  and details of our calculation will be given
elsewhere. The S-wave and P-wave short-distance coefficients have
been calculated in \cite{maltoni}, and our calculated results agree
with theirs. The D-wave short distance coefficients presented here
are new, and they are
\begin{subequations}
\begin{equation}
2\textrm{Im}f(^{3}S_{1}^{[1]})=\frac{40\alpha_{s}^{3}(\pi^{2}-9)}{81},
\end{equation}
\begin{align}
&2\textrm{Im}f(^{3}S_{1}^{[8]})=\frac{\alpha_{s}^{2}}{108}
(36N_{f}\pi+\alpha_{s}(5(-657+67\pi^2)\nonumber\\
&+N_{f}(642-20N_{f}-27\pi^{2}+72ln2)+144\beta_{0}N_{f}ln\frac{\mu}{2m_{c}})),
\end{align}
\begin{align}
&2\textrm{Im}f(^{3}P_{0}^{[8]})=\frac{5\alpha_{s}^{2}}{1296}
(648\pi+\alpha_{s}(9096-464N_{f}\nonumber\\
&+63\pi^2+2520ln2+2592\beta_{0}ln\frac{\mu}{2m_{c}}+96N_{f}ln\frac{2m_{c}}{\mu_{\Lambda}})),
\end{align}
\begin{equation}
2\textrm{Im}f(^{3}P_{1}^{[8]})=\frac{5\alpha_{s}^{3}(4107-64N_{f}-414\pi^2+48N_{f}ln\frac{2m_{c}}{\mu_{\Lambda}})}{648},
\end{equation}
\begin{align}
&2\textrm{Im}f(^{3}P_{2}^{[8]})=\frac{\alpha_{s}^{2}}{648}
(432\pi+\alpha_{s}(12561-464N_{f}\nonumber\\
&-774\pi^2+1008ln2+1728\beta_{0}ln\frac{\mu}{2m_{c}}+240N_{f}ln\frac{2m_{c}}{\mu_{\Lambda}})),
\end{align}
\begin{align}
2Imf(^{3}D_{1}^{[1]})=\frac{(321\pi^{2}-8032-29184ln\frac{\mu_{\Lambda}}{2m_{c}})\alpha_{s}^{3}}{5832},
\end{align}
\end{subequations}
where $\beta_{0}=\frac{11N_{c}-2N_{f}}{6}$, $N_c=3$, $N_f$ is the
number of flavors of light quarks. $\mu$ and $\mu_{\Lambda}$ are
renormalization and factorization scales respectively. We consider
ten processes to get the short distance coefficients in Eq.[8],
including $~gg,~ggg,~q\bar{q}$, and $~q\bar{q}g$ final states. The
contributions of $q\bar{q}$ and $q\bar{q}g$ processes are labeled by
the powers of $N_{f}$.

After calculating the short distance coefficients, we come to
determine the long-distance matrix elements. In the P-wave
charmonium decay, at leading order in $v^2$ there are two
four-fermion operators $H1$ and $H8$\cite{BBL1}, while in the case
of D-wave, there are four independent matrix elements under the
heavy-quark spin-symmetry. They are $H_{D1},H_{P8},H_{S8},H_{S1}$,
where $H_{P8}=\frac{\langle
H|\mathcal{O}_{8}({^{3}P_{0}})|H\rangle}{m_c^4}$=$ \frac{4\langle
H|\mathcal{O}_{8}({^{3}P_{1}})|H\rangle}{3m_c^{4}} =\frac{20\langle
H|\mathcal{O}_{8}({^{3}P_{2}})|H\rangle}{m_c^{4}}$, and these
relations can be derived by considering the E1 transition from
$^3D_{1}$ to $^3P_{J}$. In NRQCD, $H_{D1}$ is related to the wave
function's second derivative at the origin, while for the other
three, in the absence of lattice simulations and phenomenological
inputs, we will resort to the operator evolution equation method
suggested in \cite{BBL2}, where the authors give the result of the
matrix elements in the P-wave decay. Here we derive the following
matrix elements in the D-wave case
\begin{subequations}
\begin{equation}
H_{P8}=\frac{5}{9}\frac{8C_F}{3\beta_0}\ln(\frac{\alpha_s(\mu_{\Lambda_0})}{\alpha_{s}(\mu_{\Lambda})})H_{D1},
\end{equation}
\begin{equation}
H_{S8}=\frac{C_FB_F}{2}(\frac{8}{3\beta_0})^2\ln^2(\frac{\alpha_s(\mu_{\Lambda_0})}{\alpha_{s}(\mu_{\Lambda})})H_{D1},
\end{equation}
\begin{equation}
H_{S1}=\frac{C_F}{4N_c}(\frac{8}{3\beta_0})^2\ln^2(\frac{\alpha_s(\mu_{\Lambda_0})}{\alpha_{s}(\mu_{\Lambda})})H_{D1},
\end{equation}
\end{subequations}
where $C_F=\frac{4}{3},B_{F}=\frac{5}{12}$. We choose the region of
validity of the evolution equation: the lower limit
$\mu_{\Lambda_{0}}= m_{c}v$ and the upper limit $\mu_{\Lambda}$ of
order $m_c$.

With both the obtained short distance coefficients and long distance
matrix elements, we predict the LH decay width of $^{3}D_{1}$. The
renormalization proceeds by using the $\overline{MS}$ scheme for the
coupling constant $\alpha_s$ and the on shell scheme for the charm
quark mass. For convenience, we take the factorization scale
$\mu_{\Lambda}$ to be the same as the renormalization scale $\mu$ of
order $m_c$. We choose the pole mass $m_c=1.5\textrm{GeV}, v^2=0.3,
\mu_{\Lambda_{0}}=m_{c}v, \mu_{\Lambda}=2m_c,
\alpha_{s}(2m_c)=0.249, N_{f}=3, \Lambda_{QCD}=390\textrm{MeV},
H_{D1}=\frac{15|R''_{D}(0)|^2}{8\pi
m_{c}^6}=0.786\times10^{-3}\textrm{GeV}$\cite{BT}. At
$\mathcal{O}(\alpha^{2}_{s})$, the LH decay involves three
subprocesses $(^3P_{0})_{8}\rightarrow gg, (^3P_{2})_{8}\rightarrow
gg, (^3S_{1})_{8}\rightarrow q\bar q$, and the decay width is
estimated to be
\begin{equation}
\Gamma(^3D_1\rightarrow LH)=0.205 \textrm{MeV}.
\end{equation}
At $\mathcal{O}(\alpha^{3}_{s})$, there will be seven more
subprocesses $(^{3}S_{1})_{1,8}\rightarrow
ggg,(^{3}P_{1})_{8}\rightarrow ggg,(^{3}P_{J})_{8}\rightarrow q\bar
qg,(^{3}D_{1})_{1}\rightarrow ggg$ involved, and the result turns to
be
\begin{equation}
\Gamma(^3D_1\rightarrow LH)=0.436 \textrm{MeV}.
\end{equation}
Our result shows that in NRQCD factorization the NLO QCD correction
is even larger than the LO result. The numerical values for all
subprocesses are listed in Table 1.
If we choose $\mu_{\Lambda}=m_c,~\alpha_{s}(m_c)=0.369$, the values
of LO and NLO(the sum of LO contribution plus NLO correction) become
0.28 and 0.68 MeV respectively.
The renormalization scale $\mu$ dependence of the decay rate is
shown in Fig.(1). We see that the $\mu$-dependence at
$\mathcal{O}(\alpha^{3}_{s})$ is rather mild when $\mu>0.9m_{c}$.
For simplicity we take $\mu=2m_{c}$, where the logarithm term
$ln\frac{\mu}{2m_{c}}=0.$

\begin{table}
\caption{Subprocess decay rates of $^3D_{1}$ charmonium, where
$v^2=0.3, \mu_{\Lambda}=2m_c, \alpha_{s}(2m_{c})=0.249$.} \small
\begin{tabular}{|c|c|c|}
     \hline
     Subprocess & LO(KeV) & NLO(KeV)  \\
     \hline
      $(^3S_{1})_{1}\rightarrow LH$ & 0 &0.24\\
     \hline
      $(^3S_{1})_{8}\rightarrow LH$ & 18 &33\\
     \hline
      $(^3P_{0})_{8}\rightarrow LH$ & 184 &410\\
     \hline
      $(^3P_{1})_{8}\rightarrow LH$ & 0 &-5.8\\
     \hline
      $(^3P_{2})_{8}\rightarrow LH$ & 2.5 &4.4\\
     \hline
      $(^3D_{1})_{1}\rightarrow LH$ & 0 &-10\\
     \hline
\end{tabular}

\end{table}

\begin{figure}
\begin{center}
\includegraphics[scale=0.85]{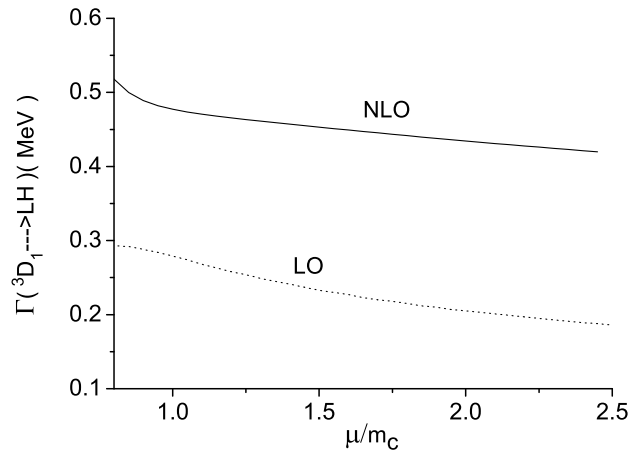}
\caption{Renormalization scale $\mu$-dependence of the decay width
of charmonium $^{3}D_{1}$ to light hadrons. Here NLO means LO
contribution+NLO correction}
\end{center}
\end{figure}

With the pole mass $m_{c}=1.5\textrm{GeV}, \alpha_{s}(2m_c)=0.249$,
$|R_{1D}''(0)|^{2}=0.015\textrm{GeV}^7$, and
$|R_{2S}(0)|^{2}=0.529\textrm{GeV}^3$, $\theta=12^{\circ}$, we find
that the three terms on the right hand side of Eq.(2) contribute
417, 5.3, 44~KeV respectively to the LH decay of $\psi(3770)$,
and result in
\begin{equation}
\Gamma(\psi(3770)\rightarrow LH)=467\textrm{KeV}.
\end{equation}
Our result shows that the D-wave LH decay is dominant, and the S-D
mixing only has a very small effect on the $\psi(3770)$ LH decay.
One important uncertainty of our prediction is associated with the
long-distance matrix elements, especially the color-octet matrix
elements. Using the same evolution equation method in $\chi_{cJ}$
decays, we find the ratio of color octet to color singlet $P$ wave
decay matrix elements agree with the lattice
calculation\cite{9609371} to within about 20\% and with the
phenomenological values\cite{Huang2,Maltoni00} to within about 30\%.
This might indicate, though not compellingly, that the uncertainty
related to the matrix elements calculated using the evolution
equation in the D-wave decays are also about (20-30)\% or (with more
confidence) less than 50\%. Other uncertainties such as the
relativistic corrections and higher order QCD radiative corrections
are beyond the scope of the present study. On the other hand,
however, we find the decay rate to be sensitive to the value of the
charm quark mass.  If we choose the pole mass
$m_{c}=1.5\pm0.1\textrm{GeV}$, $\alpha_{s}(\mu)=\alpha_{s}(2m_c)$,
and fix other parameters as before, then our prediction becomes
\begin{equation}
\Gamma(\psi(3770)\rightarrow
LH)=467^{-187}_{+338}\textrm{KeV}(\pm50\%),
\end{equation}
\begin{equation}
\textrm{Br}(\psi(3770)\rightarrow
LH)=(2.0^{-0.80}_{+1.50})\%(\pm50\%).
\end{equation}
For a small mass $m_c=1.4$~GeV, the LH decay width and branching
ratio of $\psi(3770)$ can reach 805~KeV$(\pm 50\%)$ and 3.5\%$(\pm
50\%)$ respectively, and this could be viewed as the maximum value
for the LH decay of $\psi(3770)$ in our estimation based on the
calculation at leading order in $v^2$ and next-to leading order in
$\alpha_{s}$ in NRQCD.

Together with the partial decay width of 350-400~KeV observed for
hadronic transitions and E1 transitions of the $\psi(3770)$, the
predicted annihilation (LH) decay width in Eq.(13)
will make the total non-$D\bar D$ decay width of $\psi(3770)$ to be
about 820-870~KeV for $m_c=1.5$~GeV, and 1.15-1.20~MeV for
$m_c=1.4$~GeV. The latter may be viewed as the maximum value
obtained in our approach for the total non-$D\bar D$ decay width,
corresponding to a branching ratio of about 5\% of the $\psi(3770)$
decay.

In summary, we have given a rigorous theoretical prediction for the
LH decay of $\psi(3770)$, based on NRQCD factorization at NLO in
$\alpha_{s}$ and LO in $v^{2}$. By introducing the color-octet
contributions, the results are free from infrared divergences. We
find that for the $\psi(3770)$ the D-wave contribution is dominant,
and the effect of S-D mixing is very small. Numerically, our results
do not favor either of the two experimental results measured by BES
and CLEO collaborations. We hope our theoretical result can serve as
a clue to clarify the long-standing puzzle of the $\psi(3770)$
non-$D\bar D$ decay. We urge doing more precise measurements on both
inclusive and exclusive non-$D\bar D$ decays of $\psi(3770)$ in the
future. If their total branching ratio can be as large as 10\%, it
will be a real challenge to our current understanding of QCD, and
new decay mechanisms have to be considered.

This work was supported in part by the National Natural Science
Foundation of China (No 10675003, No 10721063).

\end{document}